# Deep learning-based prediction of response to HER2-targeted neoadjuvant chemotherapy from pre-treatment dynamic breast MRI: A multi-institutional validation study

Nathaniel Braman[1*], Mohammed El Adoui[2*], Manasa Vulchi[3], Paulette Turk[3], Maryam Etesami[4], Pingfu Fu[1], Kaustav Bera[1], Stylianos Drisis[5], Vinay Varadan[1], Donna Plecha[6], Mohammed Benjelloun[2], Jame Abraham[3] and Anant Madabhushi[1,7]

[1]Case Western Reserve University, Cleveland, OH;
[2]University of Mons, Mons,, Belgium
[3]Cleveland Clinic, Cleveland, OH;
[4]Yale School of Medicine, New Haven, CT;
[5]Institut Jules Bordet, Brussels, Belgium
[6]University Hospitals Cleveland Medical Center, Cleveland, OH;
[7]Louis Stokes Cleveland VA Medical Center, Cleveland, OH
*indicates equal contribution

## Abstract

Predicting response to neoadjuvant therapy is a vexing challenge in breast cancer. In this study, we evaluate the ability of deep learning to predict response to HER2-targeted neo-adjuvant chemotherapy (NAC) from pre-treatment dynamic contrast-enhanced (DCE) MRI acquired prior to treatment. In a retrospective study encompassing DCE-MRI data from a total of 157 HER2+ breast cancer patients from 5 institutions, we developed and validated a deep learning approach for predicting pathological complete response (pCR) to HER2-targeted NAC prior to treatment. 100 patients who received HER2-targeted neoadjuvant chemotherapy at a single institution were used to train (n=85) and tune (n=15) a convolutional neural network (CNN) to predict pCR. A multi-input CNN leveraging both pre-contrast and late post-contrast DCE-MRI acquisitions was identified to achieve optimal response prediction within the validation set (AUC=0.93). This model was then tested on two independent testing cohorts with pre-treatment DCE-MRI data: 28 patients who received HER2-targeted NAC at a second institution and a 29-patient clinical trial dataset with imaging data from 3 institutions. This model achieved strong performance in a 28 patient testing set from a second institution (AUC=0.85, 95% CI 0.67-1.0, p=.0008) and a 29 patient multicenter trial including data from 3 additional institutions (AUC=0.77, 95% CI 0.58-0.97, p=0.006). Deep learning-based response prediction model was further found to exceed both a multivariable model incorporating predictive clinical variables (AUC < .65 in testing cohorts) and a model utilizing semi-quantitative DCE-MRI pharmacokinetic measurements (AUC < .60 in testing cohorts) in performance and robustness. The results presented in this work across multiple sites suggest that with further validation deep learning could provide an effective and reliable tool to guide targeted therapy in breast cancer, thus reducing overtreatment among HER2+ patients.

## 1. Introduction

Neoadjuvant chemotherapy (NAC), the administration of chemotherapy and other agents prior to surgery, constitutes the first avenue of intervention for an expanding portion of breast cancer patients (1). Given the physical (2) and financial (3) burden of treatment, it is desirable to identify which patients will respond to NAC as early as possible. However, there is currently a lack of clinically validated pre-treatment predictors of response in the neo-adjuvant setting. Predictive biomarkers of NAC response would provide immense clinical value by enabling the identification of patients who will benefit most from neoadjuvant intervention and help guide the choice of most effective NAC strategy from the inception of treatment.

There has been increased interest in the use of analytic approaches to predict therapeutic response from standard clinical imaging, most frequently dynamic contrast-enhanced (DCE) MRI in the neoadjuvant setting. Radiomics-based approaches, involved the high-throughput extraction and analysis of quantitative imaging features describing tumor texture and shape, has shown promise in assessment of response prior to (4–7) and during (8) treatment. More recently, deep learning has emerged as a promising tool for response prediction. Whereas radiomics approaches rely on explicitly defined algorithmic descriptors of imaging phenotypes, deep learning utilizes the training of a neural network to discover novel patterns best suited to perform tasks such as classification. Accordingly, studies entailing the training of a convolutional neural network (CNN), a type of neural network for the discovery of visual patterns in images, from DCE-MRI data have shown deep learning to be a powerful tool for breast cancer diagnosis (9–13), subtype classification (14), and diagnosis of metastasis (15) on DCE-MRI. Recently, some studies have shown the ability of deep learning approaches in predicting response before (16,17) and early (18,19) in NAC.

Although promising, most previous studies (20) have explored response prediction among biologically and clinically heterogeneous breast cancers. Breast cancer is frequently clinically stratified into categories based on expression of hormone receptors (21), e.g. human epidermal receptor 2-positive (HER2+), hormone receptor-positive (HR+), and triple negative (TN). While many studies have grouped these tumor categories when exploring imaging markers of NAC response, there is likely substantial value in explicitly accounting for receptor status category in this context, given that responsiveness and standard-of-care treatment strategy varies considerably among these categories (3,22). For instance, HER2+ patients will receive HER2-targeted agents such as trastuzumab and/or pertuzumab in addition to standard chemotherapy agents, whereas triple negative patients will receive only chemotherapy. Several studies (4,6) have shown that tailoring computational imaging approaches to receptor status categories improves capability to predict response, and the value of category-specific response prediction has been shown to be higher in HER2+ breast cancers (23).

Despite evidence regarding the benefits of considering subtype and therapeutic approach (4,6,23), deep learning studies to date (16–19) have pooled breast cancers across receptor status categories with varying neoadjuvant intervention approaches. Thus, the value of deep learning to predict response to a specific treatment strategy, such as HER2-targeted NAC, remains unexplored. Additionally, prior studies have both trained and tested a CNN with data from only a single institution (17) or within a clinical trial dataset with standardized acquisition protocol (16). However, deep learning faces challenges in generalizing to new sample populations and coping with unfamiliar sources of heterogeneity (24–26), potentially limiting the performance of deep learning models when evaluated within new institutions (25,27).

In this work, we utilized a multiphase CNN for the prediction of response to HER2-targeted NAC from pre-treatment DCE-MRI. Unlike all previous deep learning studies and a majority of radiomics studies, we explored breast NAC response prediction among a cohort of patients who receive a specific targeted treatment strategy: HER2+. We developed a deep learning model to identify pCR to HER2-targeted NAC from a cohort of pre-treatment DCE-MRIs acquired at a single institution. In the design and training of our model, we place an emphasis on maximizing generalizability through a lightweight, multi-input convolutional network architecture design tailored to temporal DCE-MRI data, as well as through preprocessing and training strategies to reduce the impact of variability of MRI acquisition protocol. We then tested our model on the most diverse, external multi-site dataset to date for validation of a DL approach for predicting response to NAC: comprised of patients from a second, external institution and from a clinical trial dataset collected across 3 institutions. Its performance was compared against models incorporating response-associated clinical variables and DCE-MRI semi-quantitative pharmacokinetics.

## 2. Results

### 2.1. Patient Characteristics

We investigated a total of 157 patients with HER2+ breast cancer from 5 institutions who received DCE-MRI exams prior to targeted NAC. Pathologist assessment of excised surgical samples revealed that 76 patients achieved pCR, defined as the absence of invasive cancer within the breast and axillary lymph nodes (ypT0N0/is), while 87 had residual disease following NAC (non-pCR). Patients were divided for model development and testing based on institution, in order to allow assessment for generalizability across institutions. The training (D1, n=85) and internal validation (D2, n=15) cohorts – used to learn and optimize predictive models, respectively – were formed using patients from Institution 1. A first external testing set (D3, n=28) was comprised of patients from institution 2. A second testing set (D4, n=29) consisted of patients imaged and treated at Institution 3, Institution 4, or Institution 5 as part of the BrUOG 211B multicenter clinical trial (28–30). Patients in D1, D2, and the majority of patients in D3 received a targeted NAC regimen of docetaxel, carboplatin, trastuzumab, and pertuzumab (DCTP). Patients in D4, and five patients from D3, received a regimen of only docetaxel, carboplatin, and trastuzumab (DCT).

**Table 1**. Clinical variables for the DCE-MRI datasets utilized. ER, estrogen receptor. PR, progesterone receptor. pCR, pathological complete response.

| | D1 (n=85) | | | D2 (n=15) | | | D3 (n=28) | | | D4 (n=29) | | |
|---|---|---|---|---|---|---|---|---|---|---|---|---|
| | pCR (n=42) | Non-pCR (n=43) | P | pCR (n=8) | Non-pCR (n=7) | P | pCR (n=16) | Non-pCR (n=12) | P | pCR (n=10) | Non-pCR (n=19) | P |
| Longest diameter, mm | 4.1 [1.5-10.5] | 4.7 [1.6-16.4] | 0.33 | 3.7 [1.2-8.2] | 4.2 [1.8-8.5] | 0.69 | 5.3 [1.3-12.2] | 4.2 [2.1-11.0] | 0.35 | | | |
| Mean age, years | 51.4 [28-77] | 48.8 [28-77] | 0.30 | 50.5 [38-76] | 58.4 [38-73] | 0.20 | 47.9 [31-73] | 47.4 [23-65] | 0.93 | 46.2 [32-63] | 51.3 [42-68] | 0.12 |
| ER Status (%) | | | 0.0005 | | | 0.07 | | | | | | 0.68 |
| Negative | 23 | 8 | | 3 | 0 | | 8 | 4 | 0.38 | 5 | 8 | |
| Positive | 19 | 35 | | 5 | 7 | | 8 | 8 | | 5 | 11 | |
| PR Status (%) | | | 0.007 | | | 0.45 | | | 0.23 | | | 0.37 |
| Negative | 27 | 15 | | 5 | 3 | | 9 | 4 | | 7 | 10 | |
| Positive | 15 | 28 | | 3 | 4 | | 7 | 8 | | 3 | 9 | |
| Clinical Stage | | | 0.65 | | | 0.63 | | | 0.15 | | | 0.70 |
| I | 4 | 2 | | 1 | 0 | | 1 | 3 | | 0 | 0 | |
| II | 29 | 30 | | 6 | 6 | | 9 | 7 | | 6 | 10 | |
| III | 9 | 11 | | 1 | 1 | | 6 | 1 | | 4 | 9 | |
| IV | 0 | 0 | | 0 | 0 | | 0 | 1 | | 0 | 0 | |
| Lymph Node Status | | | 0.45 | | | 0.05 | | | | | | |
| Negative | 22 | 26 | | 3 | 1 | | N/A | N/A | | N/A | N/A | |
| Positive | 20 | 17 | | 5 | 6 | | N/A | N/A | | N/A | N/A | |

DCE-MRI exams consisted of one pre-contrast t1-weighted acquisition acquired prior to and then 3-6 acquisitions following injection of a gadolinium-based contrast agent. 148 patients were imaged with a 1.5 Tesla (T) MRI scanner and 9 patients were imaged with a 3 T scanner.

## 2.2. Predictive value of clinical variable and DCE-MRI pharmacokinetic models

We evaluated the association of clinical variables available for all datasets (age, estrogen receptor (ER) status, progesterone receptor (PR) status, stage, and tumor size) and pCR in the training set. Of these, ER status (p=.0005) and PR status (p=0.007) were found to be individually significant and thus were incorporated into a multivariable clinical logistic regression model. This model identified pCR with AUC = 0.679 (95% CI: 0.45-0.87, p=0.19) within D2 (Figure 1a), but performance dropped when applied to different institutions AUC = 0.62 (95% CI: 0.44-0.77, p=0.26) for D3 (Figure 1b) and AUC=0.57 (95% CI: 0.40-0.73, p=0.50) for D4 (Figure 1c). A second model incorporating all clinical variables regardless of individual significance exhibited greater drop off between institutions (AUC=0.66 for D2, AUC = 0.54 for D3, and AUC=0.53 for D4).

We also evaluated whether a set of commonly evaluated semi-quantitative pharmacokinetic parameters, measuring temporal change in contrast enhancement on DCE-MRI, could help predict response to HER2-targeted NAC on the subset of patients with available DCE-MRI timing info (D1, n=61; D2, n=13; D3, n=22; D4, n=16). A set of 48 statistics of 8 semi-quantitative pharmacokinetic parameters describing the tumor enhancement profile voxelwise and across the entire tumor were computed on DCE-MRI with manual annotations of the tumor boundary. Within D1 and D2, permutations of seven types of classifiers paired with three to eight PK features chosen by 6 feature selection algorithms were trained to predict pCR using this set of PK features. Of these, a linear discriminant analysis (LDA) classifier with 3 features chosen by t-test was selected as the optimal pharmacokinetic model (AUC=0.81, 95% CI: 0.55-1.0 in D2) and significantly distinguished pCR in D2 (p=.025). However, this model performed poorly in external testing, achieving an AUC of 0.52 (95% CI: 0.29-0.74, p=0.87) in D3 and 0.58 (95% CI: 0.22-0.91, p=0.70) in D4 (Figure 1).

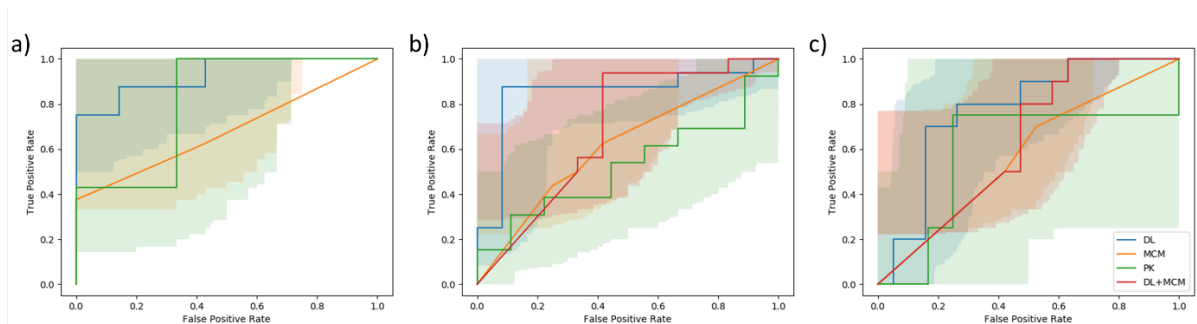

**Figure 1.** Receiver operating characteristic curves for the deep learning (DL) and comparison models (MCM: multivariable clinical model, PK: pharmacokinetic model, DL+MCM: model combining clinical variables and DL predictions) in internal validation cohort D2 (a) and external testing cohorts D3 (b) and D4 (c).

### 2.3. Performance of Deep Learning Model

A multi-input network was designed to exploit the dynamic nature of DCE-MRI by separately learning discriminative features particular to each temporal phase of DCE-MRI acquisition (Figure 2). The network was comprised of up to four distinct branches of convolutional layers, which were trained to extract patterns of response within a 2D image of a particular phase of DCE-MRI. Representations from each phase were then aggregated into a set of deep features, then these deep features were processed with a final set of dense operations summarizing the relationships between patterns at each phase.

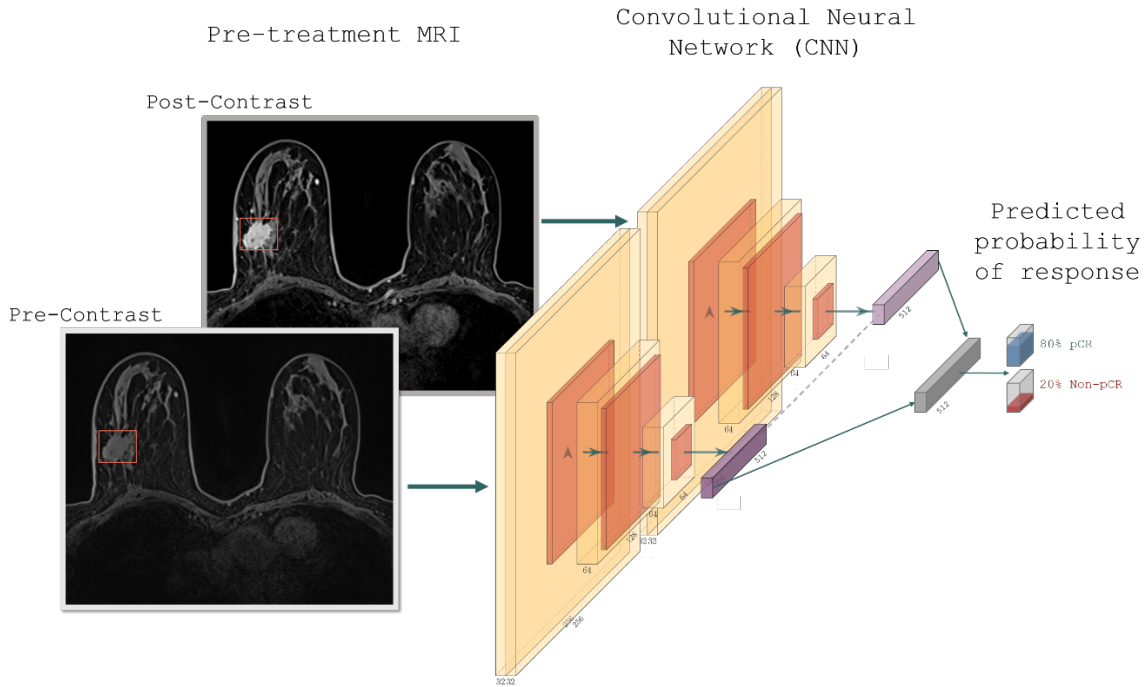

***Figure 2.*** *Convolutional Neural Network (CNN) architecture for the prediction of pathological complete response (pCR) from two DCE-MRI phases. The pre-contrast phase and a post-contrast phase were passed separately as inputs to their own convolutional branch of the two-input network. Each phase was operated on by a unique set of convolution-based operations and summarized into a set of deep features, which were then combined and processed to yield a final score indicating probability of response.*

Multiple input configurations of this deep model were explored to identify what temporal information was most informative in the prediction of pCR (Table 2). Among pairwise combinations of DCE-MRI phases, response was best predicted by a network incorporating the pre-contrast and 3$^{rd}$ post-contrast scans (Pre-3). This model achieved an AUC of 0.93 (p=.001, 95% CI=0.80-1.0) within the internal validation set D2 and was thus chosen as the final response prediction model. At a classification threshold of .5, accuracy was 86.7%, with a corresponding sensitivity of 75% and specificity of 100%. A four-input model incorporating all DCE-MRI phases (depicted in Supplementary Figure 1) was also trained, but underperformed (Accuracy=87%, AUC=0.80) relative to the best two-input model, potentially due to increased parameterization within a training dataset of limited size.

**Table 2**: AUC, sensitivity and specificity for each CNN input configuration within the internal validation dataset. *Pre,* pre-contrast phase. *1-3,* first through third post-contrast phases.

| DCE-MRI Phase Inputs | AUC | Sensitivity (%) | Specificity (%) |
|---|---|---|---|
| Pre-1 | 0.77 | 84 | 68 |
| Pre-2 | 0.81 | 86 | 79 |
| Pre-3 | 0.93 | 95 | 88 |
| 1-2 | 0.81 | 87 | 71 |
| 1-3 | 0.83 | 90 | 72 |
| 2-3 | 0.76 | 83 | 68 |
| Pre, 1, 2, and 3 | 0.80 | 90 | 72 |

The optimal two-phase model (Pre-3) was applied to the external testing cohorts. Within D3, the model significantly identified pCR with an AUC of 0.85 (p<.0001, 95% CI 0.70-1.0). At the operating point, accuracy was 86%, with a corresponding sensitivity of 81% and specificity of 92%. Likewise, discrimination of response to NAC was significant in the multi-institutional D4 cohort, achieving an AUC of 0.77 (p=0.003, 95% CI 0.61-0.91), accuracy of 79%, sensitivity of 70%, and specificity of 84%.

To provide greater interpretability to the model, we utilized gradient class activation maps (Grad-CAM) with guided backpropogation (31) to identify those image features that saliently contributed to successful predictions of therapeutic outcomes (figure 3). Intriguingly, the activation maps seemed to emphasize the tumor margins, non-mass enhancement patterns, and the surrounding peri-tumoral tissue. This finding is consistent with previous radiomic findings showing the importance of the peri-tumoral region in radiomics studies exploring NAC response prediction (4,23). Supplementary figure 2 depicts activation maps for a scan where CNN confidence was low, with differing activation between pre- and post-contrast scans.

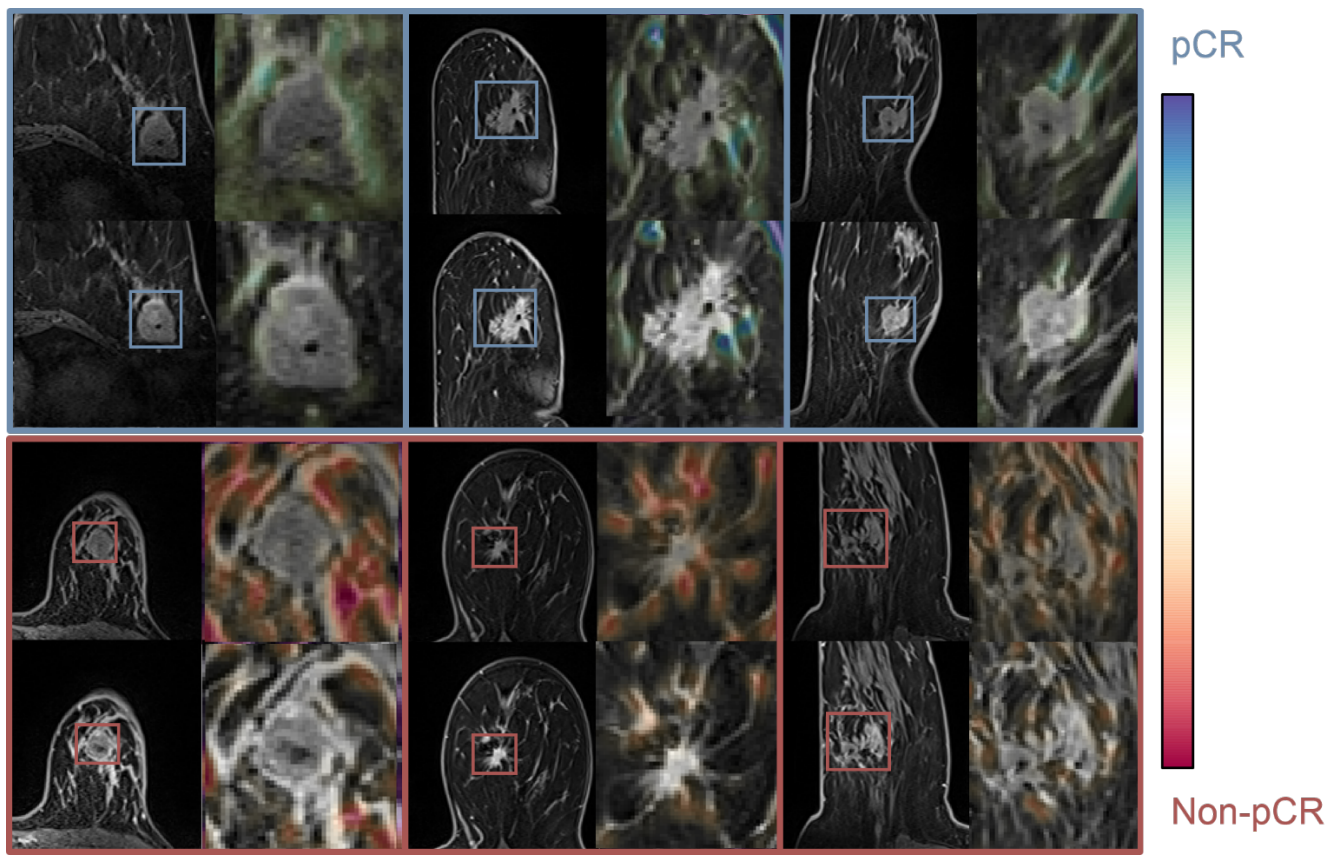

***Figure 3.*** *Model interpretability results leveraging guided Grad-CAM to identify regions contributing predictions of response (top, blue) and non-response (bottom, red). Activation maps emphasize irregular margins, heterogeneous enhancement, and the peri-tumoral region as contributing to response predictions.*

A subset analysis was performed to determine whether the model could provide predictive value complementary to IHC-based HER2+ molecular subtypes, coarsely associated with response to HER2-targeted NAC. Among ER/PR negative patients, who are hormone receptor negative and typically have elevated response to targeted therapy, the CNN successfully identified 3 of 3 patients in D3 and 6 of 8 patients in D4 who did not achieve pCR. Likewise, among the less-responsive ER/PR Positive subtype of cancers, the network identified 7 of 9 patients and 3 of 5 patients who achieved pCR in D3 and D4, respectively. Furthermore, the probabilistic output of the CNN differed significantly between pCR and non-pCR within both subtypes in D3 (ER/PR Negative, n=10, p=0.001; ER/PR Positive, n=18, p = 0.010) and ER/PR Negative in D4 (ER/PR Negative, n=13, p=0.048; ER/PR Positive, n=16, p = 0.070) by two-sided t-test. Full subset level accuracies within the testing sets are included in Supplementary Table 1.

We additionally assessed the impact of various MRI acquisition parameter on our deep learning model. Although the number of 3 T scans in the testing cohorts was limited (8 total across both cohorts), subset analysis by magnetic strength revealed that this minority group was not less accurately identified than patients who received a 1.5 T scan (Supplementary Table 2). Specifically, accuracy among patients receiving 3 T MRI was 80.0% in D3 (83% for 1.5 T) and 100% for D4 (77% for 1.5 T). We also explored the association between CNN-predicted response probability and a number of continuous acquisition parameter values (pixel spacing, slice thickness, TR, TE, number of phase encoding steps, percent sampling, flip angle, and contrast agent bolus volume) across both testing cohorts for the subset of scans with these values included in their DICOM file meta-information (Table 3). All MRI acquisition parameters were found to not be significantly correlated with CNN predictions in both D3 and D4 (p>.05), with the exception of the number of phase encoding step, which had a significant association in D4 ($r=0.45$, p=0.041, N=21), but not D3 ($r=0.29$, p=0.147, N=27).

**Table 3.** Spearman correlation between deep learning predictions and various DCE-MRI acquisition parameters in testing sets for all patients with each parameter available in DICOM metadata.

|  | D3 (n=28) | | | D4 (n=29) | | |
| --- | --- | --- | --- | --- | --- | --- |
|  | N | r | P | N | r | P |
| Pixel Spacing | 27 | 0.1753 | 0.3817 | 29 | -0.1980 | 0.3031 |
| Slice Thickness | 27 | 0.0420 | 0.8353 | 29 | 0.2964 | 0.1185 |
| Repetition Time (TR) | 27 | 0.1911 | 0.3397 | 29 | 0.0142 | 0.9418 |
| Echo Time (TE) | 27 | -0.0310 | 0.8782 | 29 | 0.1650 | 0.3923 |
| Number Of Phase Encoding Steps | 27 | 0.2865 | 0.1474 | 21 | 0.4481 | **0.0417** |
| Percent Sampling | 27 | 0.0505 | 0.8025 | 29 | 0.0942 | 0.6268 |
| Flip Angle | 27 | 0.1469 | 0.4646 | 29 | -0.0025 | 0.9897 |
| Contrast Bolus Volume | 22 | -0.1902 | 0.3965 | 28 | -0.0673 | 0.7335 |
| Time since contrast agent injection, s | 22 | 0.0113 | 0.9602 | 16 | 0.0860 | 0.7516 |

When a multivariate model combining prediction of response from the deep learning model with discriminative clinical variables (ER and PR status) was trained based on CNN output scores within D2, performance was AUC=0.69 in D3 (95% CI: 0.50-0.86, p=0.085) and AUC=0.63 in D4 (95% CI: 0.47-0.78, p=0.18). Notably, predictions from the deep learning model were found to offer significant independent value in identifying pCR when accounting for other clinical variables. The deep learning model was found to be independently associated with pCR in D3 (p=0.002) and D4 (p=0.020) when evaluated in a multivariable setting, whereas included clinical variables were not (ER status: p=0.82 in D3 and p=0.55 in D4, PR status: p=0.86 in D3 and p=0.68 in D4).

## 3. Discussion

Neo-adjuvant therapy can substantially improve survival, enable the use of breast-conserving surgery, and spare lymph node resection. Specifically, in the context of HER2-positive breast cancer, the advent of HER2-targeted therapeutic regimens has drastically improved outcomes. However, a considerable portion of HER2+ patients will still ultimately fail to respond to chemotherapy, imposing unnecessary suffering and financial burden while delaying effective intervention. Thus, there is an unmet clinical need for predictive markers that can identify prior to treatment which patients will benefit from HER2-targeted NAC.

A number of recent approaches (7,16,17,19,20,32) involving the use of both deep learning and radiomics for response prediction of NAC treated breast cancer patients have been evaluated. Whereas radiomics entails pre-defining a list of potentially predictive image features to be used in combination with a machine learning model, deep learning strategies leverage neural networks to determine the most predictive image representations (typically derived as a series of image convolutions) in order to optimally discriminate therapeutic outcomes. Ha et al. (17) trained and tested a VGG 16 (33) CNN for NAC response prediction from images of the first post-contrast phase of DCE-MRI exams, collected at a single institution. Ravichandran et al. (16) trained an AlexNet (34) CNN for response prediction within the ISPY1 clinical trial dataset, placing multiple DCE-MRI phases within the channels reserved for RGB colors in the original model. Huynh et al. (19) utilized a pre-trained VGG CNN to extract features from a multi-channel input from a 64 patient single institution dataset. El Adoui et al. utilized a multi-input CNN to predict response based on pre-treatment and during-treatment MRI data acquired at a single institution. Each of these deep learning studies utilize cohorts containing heterogeneous receptor status categories and therapeutic strategies. Apart from deep learning based approaches, several recent studies have explored response prediction through radiomics approaches at the pre-treatment time point, which leverage high-dimensional lists of pre-defined quantitative imaging features (4,6,7,35). Some of these radiomic-based studies (4,6) have shown that response prediction was improved when tailoring imaging signatures to subtype groups, suggesting that the imaging phenotype associated with response might vary with tumor biology and treatment approach. Specifically, among HER2+ breast cancers who received targeted therapy, a specialized radiomic signature was found to distinguish response-associated molecular subtypes and subsequently predict response (23). Conversely, a recent large radiomics study found that a subtype-agnostic radiomic response signature performed more poorly among HER2+ patients as compared to other subtypes (7).

In this study, we sought to develop a generalizable deep learning approach for the non-invasive prediction of benefit of a targeted neoadjuvant therapeutic regimen using only treatment-naïve dynamic MRI of breast cancer. We trained a convolutional neural network with MRI data from before and after injection of contrast agent from a single institution for the prediction of response to HER2-targeted NAC. Our approach performed strongly across institutions in identifying patients who would achieve a complete pathologic response. Furthermore, this deep learning model provided predictive value independent to response-associated sub-categories of HER2+ breast cancers; for instance, by reliably identifying responders among non-ER/PR Negative patients, who exhibit lower targeted NAC response rates despite being clinically HER2+.

Our study differed from previous deep learning approaches for breast cancer response prediction in the following ways. First, in contrast to previous work, which has largely explored computational and machine learning approaches for the prediction of NAC response within a general breast cancer population (7,16,17,19,20,32), our work explores the use of a CNN to predict response to targeted NAC for HER2+ breast cancer patients. Despite the demonstrated promise of deep learning for pre-treatment NAC response prediction, it remains underexplored in the context of receptor status category or targeted treatment regimen: likely due in part to the need for large datasets for deep learning model training. As compared to approaches non-specific to subtype, our approach has the potential to more directly guide therapeutic strategy by informing the likely therapeutic outcome of a specific targeted treatment approach. Second, our model was shown to be relatively robust across scans and institutions, an important consideration in clinical deployment and translation of these approaches. Thus far, no previous work exploring DL-based response prediction that we are aware of has explicitly demonstrated generalizability across data from multiple different sites not included in training. Our model was trained using data from a single institution in a manner specifically to improve its generalizability, through strategies such as a lightweight convolutional network design to mitigate overfitting and augmentation approach intended to introduce heterogeneity likely found in a multi-site setting (e.g. variations of spatial resolution). We then validated our predictive model on pre-treatment MRI data from a second institution and a multi-center clinical trial dataset and found performance to perform consistently across scanners and imaging data from four new institutions. As such, this is the most rigorous validation of a deep learning approach for breast cancer response prediction from imaging to date. Furthermore, we thoroughly interrogated the association of DL model predictions with magnetic strengths (Supplementary Table 2) and other MRI settings (Table 3), and found our model to be largely independent of acquisition parameters across both cohorts. Third, the deep learning strategy presented was tailored to the dynamic nature of DCE-MRI through a multi-input CNN architecture that can accept and process multiple temporal contrast phases in parallel. Previous approaches have restructured DCE-MRI data to accommodate computer vision techniques designed for natural images; for instance, by aggregating DCE-MRI phases into an image's RGB channels of an input image (16,19) or by utilizing only one DCE-MRI temporal phase (17). However, such approaches might be limited in their ability to capture prognostic differences in tumor presentation throughout contrast enhancement (36,37) during a dynamic MRI exam. The multi-input network employed in our study learns discriminative features that are specific to each phase of DCE-MRI acquisition through separate convolutional branches to extract response-associated image patterns particular to the pre- or post-contrast presentations. The dense operations following these convolutional branches enable the identification of relationships between CNN-extracted representations of individual contrast phases in order to predict a patient's ultimate therapeutic response. This CNN architecture was found to offer stronger performance and generalizability compared to a semi-quantitative pharmacokinetic model of direct measures of a tumor's enhancement on DCE-MRI, previously shown to be associated with NAC outcomes and survival (37–43).

A further distinction of our approach relative to radiomics methods is the ability to discover novel patterns of response without a-priori definition of a set of quantitative image features and the spatial regions where they will be discriminative. The large majority of radiomics approaches require precise, often manually-defined contours of the tumor (44) for analysis. In contrast, deep learning can utilize courser annotations such as a box containing the tumor and discover the patterns and regions most critical to a classification task. To delve further into this advantage, we utilized gradient class activation maps (Grad-CAM) to identify the portions of images that drove the CNN's predictions of therapeutic outcomes (figure 2). The CNN honed in on the tumor's margin and the peri-tumoral region, consistent with prior findings exploring radiomics of the peritumoral region and parenchyma (4,23,45–48) for outcome prediction, particularly in HER2+ breast cancer (23). Conceivably, these attention maps could be utilized by radiologists and oncologists to better understand CNN guidance regarding targeted therapy.

Our study did have its limitations. First, the size of our dataset is limited. Our emphasis on response prediction for a specific targeted therapeutic approach restricts the size of available imaging data in the NAC setting. This constraint might explain the inferior performance of a model containing all DCE-MRI phases as compared to two-phase model, which could be over-parameterized to our limited training dataset. Nevertheless, our study is among the largest investigating deep learning-based response prediction from breast MRI (16,17,19) and the largest study to investigate response prediction to a specialized treatment strategy or within specific molecular subtype categories (4,6,20,23,49). Despite limitations posed by the size of our specialized cohort, the demonstration of strong response prediction among a number of external institutions unseen during training is strongly suggestive of its reproducibility. Second, our approach does require manual data preprocessing, in particular drawing a box indicating the tumor region-of-interest. While delineating these tumors manually would require the time of trained radiologists if clinically implemented, it is important to note that our deep learning approach's requirement of only a box annotation would be significantly less cumbersome to implement and less prone to inter-reader variability than radiomics approaches that require delineation of precise tumor boundaries. Future work aimed at automating this step, potentially through a deep learning-based region proposal approach (50), could be a worthwhile direction for improving its efficiency. Third, while a unique strength of our work is consideration of a specific therapeutic strategy, it should be noted that the exact HER2-targeted NAC regimen utilized in the validation cohorts did vary. For instance, patients in D4 did not receive pertuzumab, but did receive the HER2-targeted agent trastuzumab. While performance remained strong despite differing HER2-targeted NAC regimens, future work is needed to more fully determine the extent to which the model is generalizable across HER2-targeted NAC strategies.

In summary, we demonstrate that deep learning is a robust and effective strategy for response prediction from breast MRI before initiation of targeted therapy. By predicting outcomes to HER2-targeted NAC, our model could be feasibly utilized to more precisely target treatment among HER2+ breast cancers. Furthermore, the implications of imaging markers for response to HER2-targeted therapies has wide-reaching potential benefit beyond breast cancer. HER2-targetted therapies have increasingly shown benefit in other cancer domains, but with lower overall rates of response and, as a result, high need for predictive biomarkers (51). Future work could explore the use of a neural for response prediction in such domains, potentially through transfer learning using the current breast prediction model.

## 4. Methods and materials
### 4.1 Datasets
#### 4.1.1 Training (D1) and Internal Validation (D2)

A cohort of 100 patients received HER2-targeted NAC consisting of docetaxel, carboplatin, trastuzumab, and pertuzumab (DCTP) who received pre-treatment imaging were retrospectively identified. 50 patients achieved pCR, as defined as the complete absence of residual invasive disease within the breast and axillary lymph nodes (ypT0N0/is) on post-NAC surgical samples. The remaining 50 patients had residual disease following NAC, and were considered non-pCR.

Patients were divided randomly into an 85 patient training cohort (D1) and a 15 patient validation cohort (D2) for optimizing model architecture, inputs, and hyperparameters. Clinical variables and acquisition parameters for all cohorts are summarized in Table 1 and Supplementary Table 1, respectively.

#### 4.1.2 Testing Cohort D3

28 HER2+ patients who received DCE-MRI exams at Institution 2 between March 1$^{st}$, 2012 and May 15$^{th}$, 2016 prior to targeted NAC, a subset of a retrospective cohort previously described in (4) and (23), formed the first external testing cohort D3. Of the 28 patients, 16 achieved pCR on surgical specimen (ypT0N0/is), while 12 were non-pCR. The majority of patients received a HER2-targeted regimen of DCTP (n=23), while 5 received only docetaxel, carboplatin, and trastuzumab (DCT).

#### 4.1.3 Testing Cohort D4

29 patients obtained from the BrUOG 211B multicenter preoperative clinical trial (NCT00617942) formed the second external testing cohort D4. Patients were treated with DCT and imaged at one of three institutions between April 27, 2012, through September 4, 2015: 1) Brown University Oncology Research Group participating hospitals, Providence, Rhode Island, 2) Yale Cancer Center, New Haven CT, or 3) City of Hope Comprehensive Cancer Center, Duarte, California. 10 patients achieved pCR (ypT0N0/is).

**Table 4.** Acquisition parameters for T1-weighted for breast dynamic contrast-enhanced MRI. T, tesla; mm, millimeters; ms, milliseconds; TE, echo time; TR, repetition time

|  | **D1/D2** | **D3** | **D4** |
|---|---|---|---|
|  | *Institution 1* | *Institution 2* | *Institution 3* |
| Magnetic Strength |  |  |  |
| 1.5 T | 99 | 23 | 26 |
| 3 T | 1 | 5 | 3 |
| Spatial Resolution, mm | 0.905 [0.417-1.136] | 0.766 [0.568-1.042] | 0.712 [0.498-1.063] |
| Slice Thickness, mm | 1.169 [1.000-2.500] | 1.332 [0.900-3.000] | 1.890 [1.000-2.400] |
| Repetition Time (TR), ms | 4.978 [4.220-6.200] | 5.065 [4.430-6.926] | 5.229 [3.760-6.464] |
| Echo Time (TE), ms | 2.182 [1.590-3.200] | 1.769 [1.380-3.389] | 2.123 [1.430-3.108] |
| Number of Phase Encoding Steps | 299[144 -387] | 312 [261-352] | 294 [225-338] |
| Percent Sampling | 84.0 [80-100] | 84.5 [75-121] | 74.3 [60-100] |
| Flip Angle | 9.5 [8-15] | 10.4 [10-12] | 12.0 [10-15] |
| Contrast Agent Bolus Volume | 14.7 [7-20] | 16.2 [11-30] | 16.8 [9.7-36.0] |
| Number of Post-Contrast Acquisitons | 3.35 [3-5] | 5.05 [4-6] | 3.524 [2.000-5.000] |
| Average Time Between Acquisitions, s | 145 [67.0-211] | 86.852 [66.0-105] | 115.660 [27.800-219.752] |
| Post-Contrast Scan Time, s | 469 [268-631] | 438 [331-628] | 363.380 [120.000-659.255] |

## 4.2 Data preprocessing and preparation

In this study, pre-treatment DCE-MRI volumes with pre-contrast and 3 or more post-contrast phases were utilized to predict the response to neoadjuvant therapy. To eliminate the heterogeneous distribution of intensity throughout the breast on DCE-MR images due to inconsistent magnetic fields of the MRI machine, a bias-correction was applied to each volume. A rectangular volume of interest (VOI) containing the largest slice of the tumor was specified by an experienced radiologist with 7 years' experience practicing in breast radiology (P.T.) on post-contrast DCE-MRI subtraction images while consulting radiology and pathology reports. The rectangular region was then expanded to all neighboring slices containing tumor enhancement. Each DCE-MRI phase volume for each patient was cropped to this expanded VOI to yield 3D sub-volumes containing the tumor volume (Supplementary Figure 2). In the external validation cohorts, exact tumor boundaries were previously manually delineated on several slices by two radiologists working in consensus (D.P. 25 years practicing experience and M.E. 6 years residency and fellowship training). To be consistent with annotation protocol applied to institution 1, rectangular bounding boxes containing these annotations were derived and used in this study. Processing with a CNN requires inputs with a consistent size, thus VOIs were zero-padded to yield a consistent input size of 156 x 156 x 3 pixels (Supplementary figure 3), with 3 adjacent 2D slices included as channels of the CNN input. Each input was normalized relative to its maximum grayscale level, so that all inputs had values between 0 and 1.

## 4.3 Multi-phase Convolutional Neural Network Architecture

The model used was inspired by previous work exploring multi-input CNNs for the prediction of response based on multiple scans throughout treatment (18). In contrast to this approach, we instead utilized a similar multi-input architecture to process multiple DCE-MRI temporal phases (acquired on the same pre-treatment imaging exam date) as input, with each phase processed by a separate convolutional branch. A model utilizing two DCE-MRI phases is depicted in Figure 1.

Each phase-specific branch of the neural network consists of four blocks of 2D convolution layers, each followed by a non-linear activation function and max pooling layer (with the exception of the first block, where max pooling was not used). 32 kernels were used for each convolutional layer inside the first and second blocks for each branch, and 64 for the third and fourth blocks. A constant filter size of 3x3 pixels was used throughout the network. A dropout layer was applied after every two convolutional blocks, in order to help prevent overfitting and thus increase model generalizability. Deep features from each branch were concatenated, then processed with a fully connected layer of 512 units, followed by an activation function with dropout. A second dense layer with 2 units yields the final output of the network, a probabilistic value indicating likelihood of pCR, was obtained from a final fully connected layer with sigmoid activation. The output of this layer is given by:

$$\sigma = \frac{1}{1 + e^{-z}}$$

Where $z$ is the sum of the $m$ deep features from the penultimate layer, $x_i$, multiplied with learned weights, $w_i$, plus a learned bias term, $B$:

$$z = \sum_{i=1}^{m} w_i x_i + B$$

Multiple versions of the model with different input DCE-MRI phases were explored. A set of models that take two DCE-MRI phases as inputs were first explored. Models were trained using all possible phase combinations to identify the pair of DCE-MRI phases providing the best performance within the internal validation set. Therefore, six pairwise combinations were used to train the two-phase CNN. The full Keras summary of model layers and parameters is included as Supplementary Table 3.

Based on the same architecture, we additionally evaluated an extended architecture that considers four phases of DCE-MRI at one time. Therefore, four parallel CNNs were used. Their outputs were then concatenated and then processed by the network to yield a final classification based on all DCE-MRI phases. Supplementary Fig. 1 illustrates the architecture of this model.

### 4.4 Deep learning model training

The weights of the deep network were randomly initialized (52), then updated directly based on the 85 patient training set in an iterative fashion across 80 training epochs. Data augmentation was applied to synthetically expand the training set and improve generalizability of the model (Supplementary Figure 4). Operations including rotations, flips, and translations applied randomly on every epoch to each slice separately to maximize training data. An additional augmentation strategy, random resizing, was applied to mitigate the effects of heterogeneous voxel resolution between scans.

In addition to fine tuning the architecture and inputs of the network, it was additionally necessary to optimize the hyper-parameters that dictate how the model will be trained. A grid search was performed to assess a number of different hyper-parameter combinations, and the best configuration was determined based on performance within D2. The hyper-parameters and values assessed in this grid search, as well as the best-performing model configuration, are included in Supplementary Table 4.

The best-performing model was trained using a binary cross entropy loss function and the Stochastic Gradient Decent (SGD) (53) optimization function, with a learning rate of $5 \times 10^{-4}$. The deep neural network architecture and training was implemented in Python, using Keras 2.2.4 API with Tensorflow 1.9.0 backend (Python 3.5.2), utilizing the following hardware:
- CPU: 16 cores, 2.10 GHz clock speed, 128 GB of RAM memory
- 4 GPUs: Nvidia P100, 3584 CUDA cores, 10.6 TeraFLOPS, 16 GB of memory.

## 4.5 Semi-quantitative pharmacokinetic model

A set of 8 semi-quantitative pharmacokinetic (PK) parameters (description and formulae listed in Supplementary Table 5) were computed voxel-wise and summarized for each patient by computing their mean, median, standard deviation, skewness and kurtosis values across the tumor, yielding a total of 40 statistics characterizing the distribution of PK measurements across tumor voxels. Additionally, these 8 measures were computed for the entire tumor based on the average intra-tumoral intensity at each DCE-MRI phase. In total, 48 features pertaining to PK measurements were considered and evaluated in the comparative model.

To ensure a fair comparison against deep learning, a number of classification models were explored and optimized for response prediction from PK statistics. Within D1 via cross-validation, pharmacokinetic features were reduced to sets of three to eight top features using 6 different feature ranking strategies (t-test, Wilcoxon rank-sum test, entropy, Bhattacharyya distance (54), area under the ROC curve, minimum redundancy maximum relevance (55)) and used to train 7 different classifier models: linear discriminant analysis (LDA), quadratic discriminant analysis (QDA), diagonal LDA, diagonal QDA, logistic regression, support vector machine, and random forest of 100 trees). For each classifier type, the best performing feature set was evaluated on D2. Of these, the best performing classifier was chosen as the optimal PK model. Optimal configurations of each classifier, along with their performance in D1 and D2, are included in Supplementary Table 6.

### 4.6 Statistical Analysis

The output of the deep learning model was a 0 to 1 probabilistic score, where 1 corresponded to a high confidence prediction of pCR following NAC. Within the validation and testing cohorts, performance was assessed by area under the receiver operating characteristic curve (AUC), as well as the sensitivity, specificity, and accuracy corresponding to a threshold of .5. Significance of the AUC was computed in MedCalc statistical software (56) using the method of DeLong et al. (57) to calculate standard error. 95% confidence intervals for the ROC curve were computed by bootstrapping across 1,000 iterations.

As a comparative strategy, multivariable clinical models were trained and evaluated incorporating clinical variables common to all datasets: age, estrogen receptor (ER) status, progesterone receptor (PR) status, stage, largest tumor diameter, and lymph node positivity. Each variable was individually assessed for significant differences between pCR and non-pCR by Wilcoxon rank sum test for continuous variables (age, largest diameter) and chi-squared test for discrete variables (ER status, PR status, stage, lymph node positivity) in D1. All variables found to be individually significant were included in a multivariable logistic regression model and assessed for their ability to predict response in the D2-D4. A multivariable model incorporating both significant clinical variables and prediction from the DL model was trained in D2 (due to overly confident predictions of the DL model among training data), to evaluate the collective performance of both clinical information and CNN predictions in the testing sets. Each variable was assessed for independent significance when considering the model's other variables by t-test of its coefficient in the logistic regression model.

Conflicts of Interest

Nathaniel Braman: IBM Research - Former Employment. Anant Madabhushi: Inspirata-Stock Options/Consultant/Scientific Advisory Board Member, NIH Academic-Industry Partnership grants, Sponsored Research, Elucid Bioimaging Inc.-Stock Options, PathCore Inc-NIH Academic Industrial Partnership, Philips – Sponsored Research. Donna Plecha: Hologic Inc Research Grant. Vinay Varadan: Curis, Inc. - Sponsored Research; Philips Healthcare - Sponsored Research

**Supplementary Figure 1.** Four input CNN for prediction of NAC response from dynamic MRI. A pre-contrast and three sequential post-contrast DCE-MRI volumes are processed separated by convolutional branches. Features from each input branch are concatenated and processed to yield a final response prediction

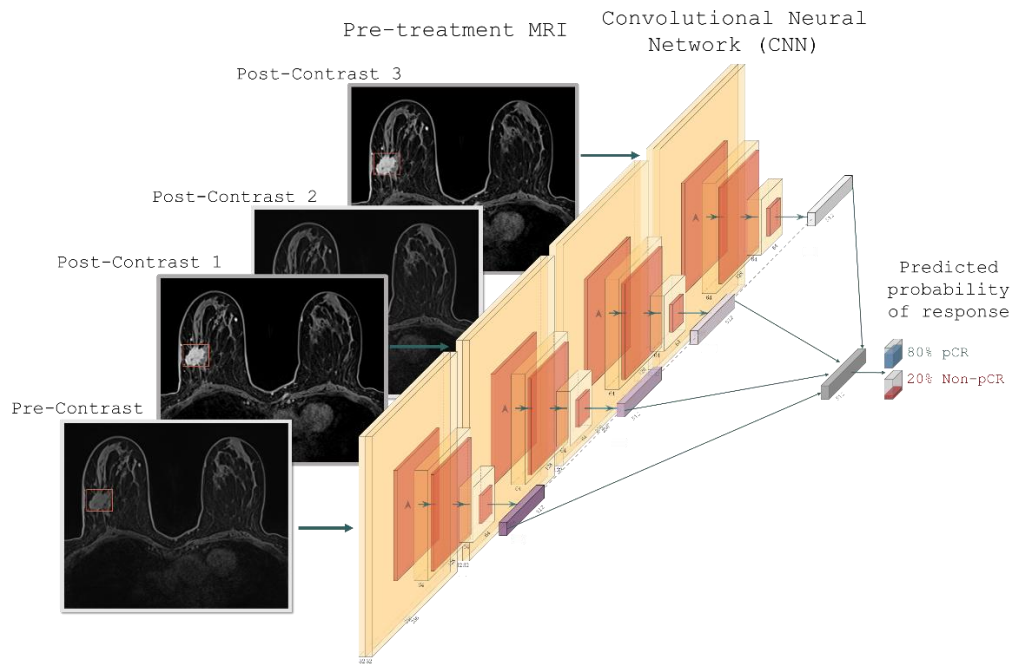

**Supplementary Figure 2.** Gradient class activation map for misclassified non-pCR patient with low confidence prediction. Class activation is split between pCR and non-pCR for pre- and post-contrast phases.

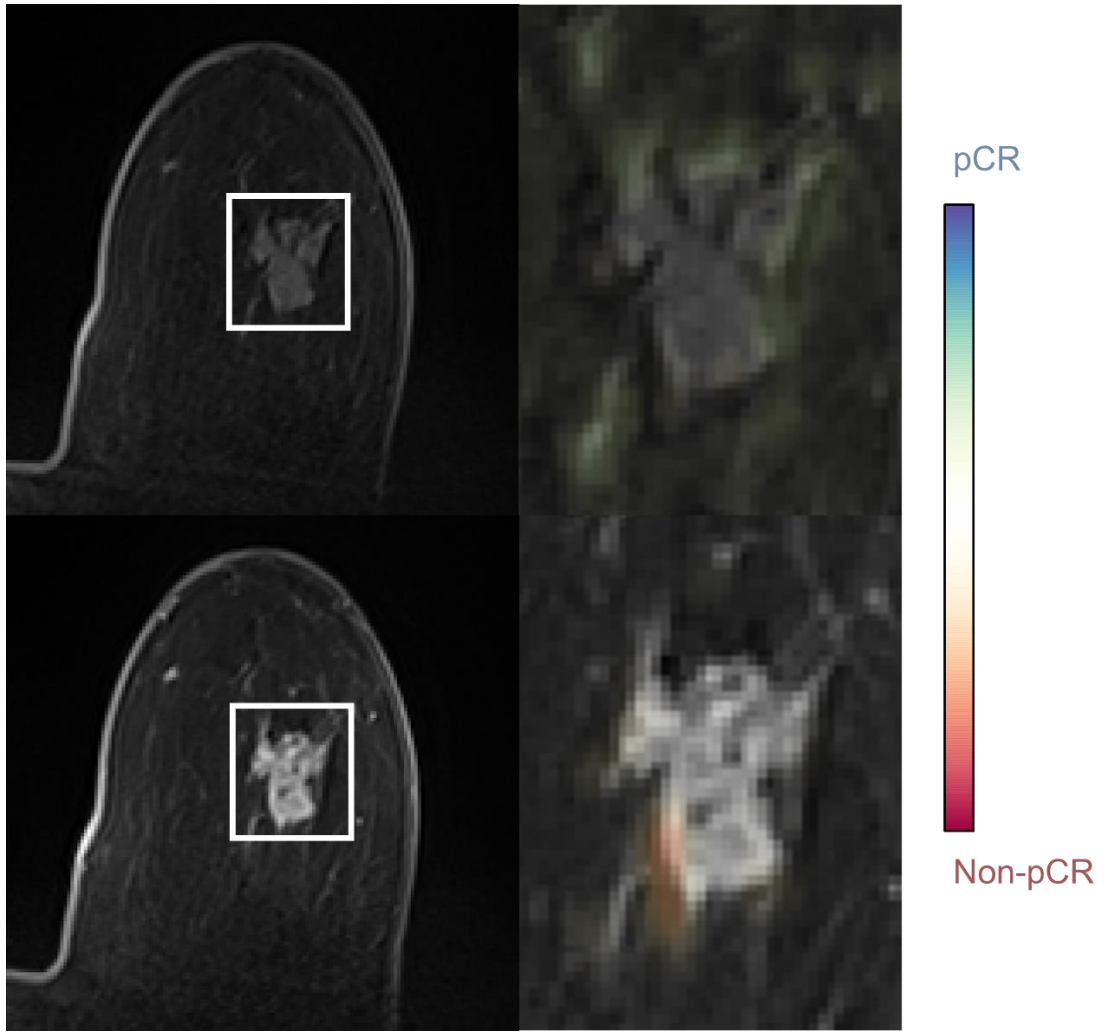

**Supplementary Table 1.** Subset analysis of performance stratified by NAC outcome and hormone receptor status in validation and testing cohorts. HR, Hormone receptor.

|  | D2 (n=15) | | D3 (n=28) | | D4 (n=29) | |
|---|---|---|---|---|---|---|
|  | **Non-pCR** | **pCR** | **Non-pCR** | **pCR** | **Non-pCR** | **pCR** |
| **HR -** | N/A | 2/3 | 3/3 | 6/7 | 6/8 | 4/5 |
| **HR +** | 7/7 | 4/5 | 8/9 | 7/9 | 10/11 | 3/5 |

**Supplementary Table 2.** Subset analysis of performance stratified by NAC outcome and scanner magnetic strength in validation and testing cohorts. T, Tesla.

|       | D3 (n=28) |       | D4 (n=29) |      |
|-------|-----------|-------|-----------|------|
|       | Non-pCR   | pCR   | Non-pCR   | pCR  |
| 1.5 T | 9/9       | 11/14 | 13/16     | 7/10 |
| 3 T   | 2/3       | 2/2   | 3/3       | N/A  |

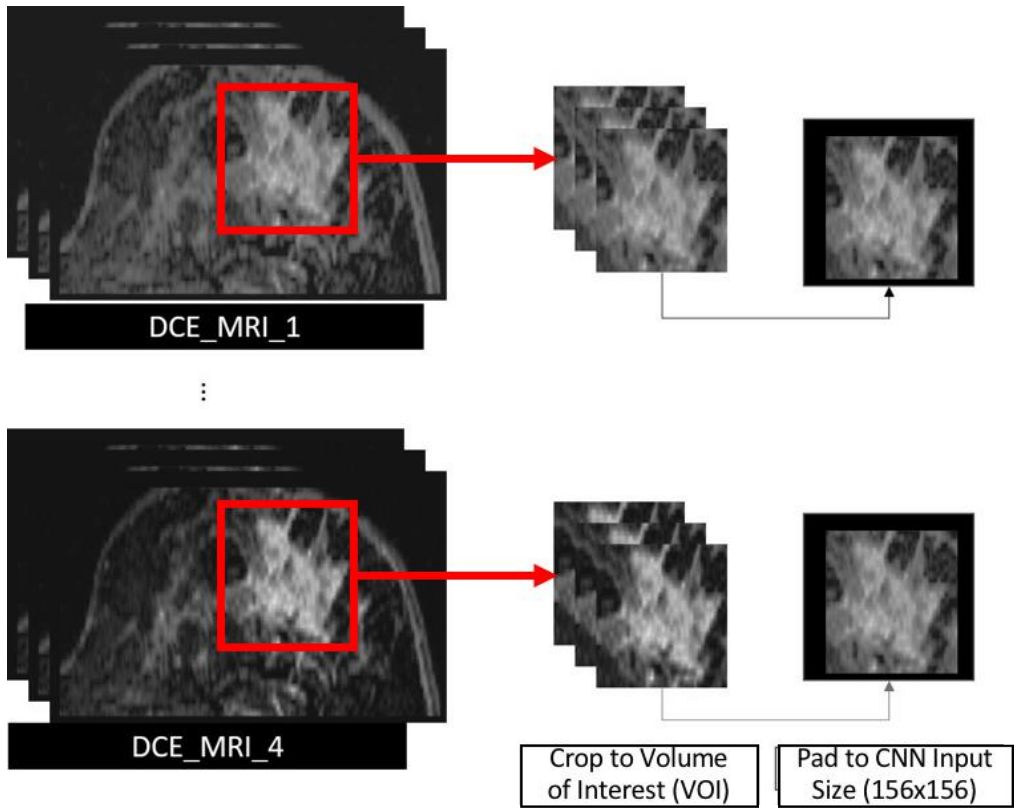

**Supplementary Figure 3:** Data preprocessing: each DCE-MRI temporal volume is cropped to the volume of interest (VOI) and zero-padded to a size of AAAxBBB, the input dimensions of the CNN.

**Supplemental Table 3.** Full network architecture and all parameters for the two input deep learning model, obtained using model.summary() in Keras.

| Layer (type) | Output Shape | Param # | Connected to |
|---|---|---|---|
| INPUT_DCEMRI_PRE (InputLayer) | (None, 156, 156, 3) | 0 | |
| INPUT_DCEMRI_POST (InputLayer) | (None, 156, 156, 3) | 0 | |
| conv2d_1 (Conv2D) | (None, 154, 154, 32) | 896 | INPUT_DCEMRI_PRE[0][0] |
| conv2d_2 (Conv2D) | (None, 154, 154, 32) | 896 | INPUT_DCEMRI_POST[0][0] |
| max_pooling2d_1 (MaxPooling2D) | (None, 77, 77, 32) | 0 | conv2d_1[0][0] |
| max_pooling2d_2 (MaxPooling2D) | (None, 77, 77, 32) | 0 | conv2d_2[0][0] |
| conv2d_3 (Conv2D) | (None, 75, 75, 32) | 9248 | max_pooling2d_1[0][0] |
| conv2d_4 (Conv2D) | (None, 75, 75, 32) | 9248 | max_pooling2d_2[0][0] |
| max_pooling2d_3 (MaxPooling2D) | (None, 37, 37, 32) | 0 | conv2d_3[0][0] |
| max_pooling2d_4 (MaxPooling2D) | (None, 37, 37, 32) | 0 | conv2d_4[0][0] |
| dropout_1 (Dropout) | (None, 37, 37, 32) | 0 | max_pooling2d_3[0][0] |
| dropout_2 (Dropout) | (None, 37, 37, 32) | 0 | max_pooling2d_4[0][0] |
| conv2d_5 (Conv2D) | (None, 35, 35, 64) | 18496 | dropout_1[0][0] |
| conv2d_6 (Conv2D) | (None, 35, 35, 64) | 18496 | dropout_2[0][0] |
| max_pooling2d_5 (MaxPooling2D) | (None, 17, 17, 64) | 0 | conv2d_5[0][0] |
| max_pooling2d_6 (MaxPooling2D) | (None, 17, 17, 64) | 0 | conv2d_6[0][0] |
| conv2d_7 (Conv2D) | (None, 15, 15, 64) | 36928 | max_pooling2d_5[0][0] |
| conv2d_8 (Conv2D) | (None, 15, 15, 64) | 36928 | max_pooling2d_6[0][0] |
| max_pooling2d_7 (MaxPooling2D) | (None, 7, 7, 64) | 0 | conv2d_7[0][0] |
| max_pooling2d_8 (MaxPooling2D) | (None, 7, 7, 64) | 0 | conv2d_8[0][0] |
| dropout_3 (Dropout) | (None, 7, 7, 64) | 0 | max_pooling2d_7[0][0] |
| dropout_4 (Dropout) | (None, 7, 7, 64) | 0 | max_pooling2d_8[0][0] |
| flatten_1 (Flatten) | (None, 3136) | 0 | dropout_3[0][0] |
| flatten_2 (Flatten) | (None, 3136) | 0 | dropout_4[0][0] |
| concatenate_1 (Concatenate) | (None, 6272) | 0 | flatten_1[0][0]<br>flatten_2[0][0] |
| dense_1 (Dense) | (None, 512) | 3211776 | concatenate_1[0][0] |
| dropout_5 (Dropout) | (None, 512) | 0 | dense_1[0][0] |
| dense_2 (Dense) | (None, 2) | 1026 | dropout_5[0][0] |

Total params: 3,343,938
Trainable params: 3,343,938
Non-trainable params: 0

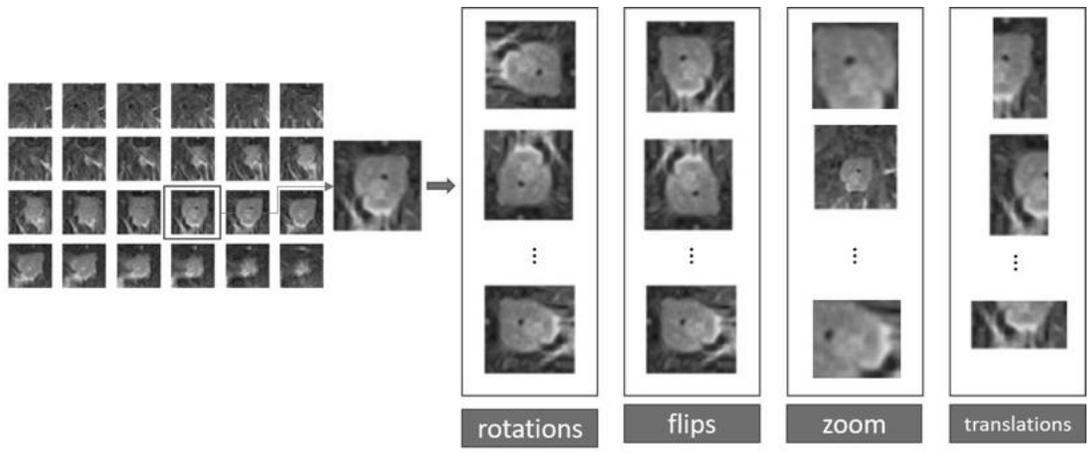

**Supplementary Figure 4:** Data augmentation based on random rotations, horizontal and vertical flips, zoom and translations

**Supplementary Table 4:** Tested and optimal values of the training parameters

| Parameter | Tested values | Optimal value |
|---|---|---|
| Learning rate (lr) | 0.05, 0.005, 0.0005, 0.00005 | 0.00005 |
| Batch size | 2, 4, 8, 16, 32 | 8 |
| Momentum rate | 0.8, 0.9, 0.99 | 0.99 |
| Weight initialization | Normal, Uniform, Glorot | Normal |
| Per-parameter adaptive learning rate methods | Stochastic gradient descent (SGD), RMSprop, Adagrad, Adadelta, Adam | SGD |
| Learning rate decay | Yes, no | Yes (1e-6 per epoch) |
| Activation function | Sigmoid, ReLU, elu | ReLU |
| Dropout rate | 0.1, 0.25, 0.3, 0.5, 0.75 | 0.25 (blocks 1 and 2) <br> 0.30 (blocks 3 and 4) <br> 0.40 (dense layer) |

**Supplementary Table 5.** Features computed for the semiquantitative pharmacokinetic model. $S_0$, $S_1$, $S_{final}$, and $S_{peak}$ are the MRI signal intensity values on the pre-contrast phase, first post-contrast phase, final post-contrast phase, and post-contrast phase of maximum intensity. $t_0$, $t_1$, $t_{final}$, and $t_{peak}$ are the corresponding times at which those acquisitions were collected.

| Variable | Description | Formula |
|---|---|---|
| Time-to-Peak | Time in seconds of scan exhibiting peak intensity due to contrast enhancement throughout DCE-MRI exam | $t_{peak} - t_0$ |
| Maximum Percent Enhancement | Percent intensity increase between max post-contrast and pre-contrast phase | $S_{peak}/S_0$ |
| Early Percent Enhancement | Percent intensity increase between first post-contrast and pre-contrast phase | $S_1/S_0$ |
| Late Percent Enhancement | Percent intensity increase between final post-contrast and pre-contrast phase | $S_{final}/S_0$ |
| Signal Enhancement Ratio | Ratio of intensity increase of first post contrast phase to peak intensity increase, relative to pre-contrast phase | $(S_1-S_0)/(S_{peak}-S_0)$ |
| Final Phase Enhancement Ratio | Ratio of intensity increase of first post contrast phase to final intensity increase, relative to pre-contrast phase | $(S_1-S_0)/(S_{final}-S_0)$ |
| Rate of Uptake | Rate of enhancement between peak enhancement and pre-contrast phase | $(S_{peak}-S_0)/(t_{peak}-t_0)$ |
| Rate of Washout | Rate of intensity decay between peak enhancement phase and final DCE-MRI acquisition | $(S_{final}-S_{peak})/(t_{final}-t_{peak})$ |

**Supplementary Table 6.** Configurations and associated performance of optimal classification models incorporating statistics of voxelwise PK measures. For each classifier, the best performing number of features and feature selection scheme was chosen based on performance in cross-validation (D1), then that configuration was applied to D2. An LDA classifier with 3 features selected by t-test was chosen as the final PK classifier.

|  | No. Features | Feature Selection Algorithm | D1 | D2 |
|---|---|---|---|---|
| **LDA** | 3 | t-test | 0.634 | 0.810 |
| **QDA** | 8 | t-test | 0.547 | 0.738 |
| **DLDA** | 3 | t-test | 0.638 | 0.762 |
| **DQDA** | 3 | t-test | 0.581 | 0.762 |
| **Logistic Regression** | 3 | wilcoxon | 0.634 | 0.667 |
| **SVM** | 3 | t-test | 0.623 | 0.762 |
| **Random Forest** | 7 | wilcoxon | 0.639 | 0.750 |